\documentstyle[preprint,aps]{revtex}  
\tightenlines

\newcommand{\intercal}{T}
\begin{document}
\draft
\preprint{\today}



\title{ Pairing-correlations and particle-number projection methods } 
\author{J.A. Sheikh$^{1}$, P. Ring$^{1}$,
E. Lopes$^{1}$ and R. Rossignoli$^{1}$}
\address{$^1$Physik-Department, Technische Universit\"at M\"unchen,
D-85747 Garching, Germany 
}

\date{\today}
\maketitle

\begin{abstract}
A systematic study of the pairing-correlations derived from various
particle-number projection methods is performed in an exactly soluble
cranked-deformed shell model Hamiltonian. It is shown that most
of the approximate particle-number projection methods including
the method of Lipkin-Nogami, which is used quite extensively in
nuclear structure studies, breakdown in the weak pairing limit.
The results obtained from the recently formulated number-projected
Hartree-Fock-Bogoliubov (PHFB) equations, on the other hand, are in complete
agreement with the exact solutions of the model Hamiltonian. The pairing-energy
calculated from PHFB method is shown to be finite in all the studied limits.
More importantly, the numerical work involved in the solution of the PHFB equations
appears to be comparable to the solution of the bare HFB equations.
\end{abstract}

\pacs{PACS numbers : 21.60.Cs, 21.10.Hw, 21.10.Ky, 27.50.+e}

\section{Introduction}

Pairing-correlations play a central role in describing the
properties of atomic nuclei \cite{bmii,rs80}. The bulk of the
nuclear phenomena, for instance, the odd-even mass differences
and moments of inertia of deformed nuclei can be described by
considering nucleons to be in a superfluid paired phase. The
recent discovery of halo nuclei have also elucidated the
importance of pairing-correlations in the formation of such
structures \cite{be91,mr96,pvl97}. In the traditional nuclear structure
models, the pairing-correlations are treated in
Bardeen-Cooper-Schrieffer (BCS) or Hartree-Fock-Bogoliubov (HFB)
approximations \cite{rs80}. These approaches use a simple ansatz
of a product-state of quasiparticles for the ground-state
wavefunction. The product ansatz leads to a set of non-linear
equations in the BCS theory or to the diagonalization of a matrix
in the case of HFB approach. There exist now very fast numerical
methods to find accurate solutions of these problems. The major
advantage in the mean-field models is that the number of
non-linear equations or the dimensionality of the matrix to be
diagonalised is quite small as compared to the exact analysis,
which is impossible to apply in a realistic configuration space of
medium and heavy mass nuclei.

Nevertheless, the product ansatz of the mean-field approach breaks
the symmetries which the orginal many-body Hamiltonian obeys. For
instance, in the case of BCS theory, the product-state violates
the gauge-symmetery associated with the particle-number. The BCS
product-state of quasiparticles does not have a well defined
particle-number. The most undesirable feature associated with the
broken symmetry is that it leads to sharp phase transitions. These
phase transitions may be reasonable for a system with a very large
number of particles, for example, a metallic superconductor. But
for a finite mesoscopic system these phase transitions are washed
out by fluctuations and are not observed in the experimental data
or in the exact study of toy models. The sharp transition is an
artifact of the simple ansatz of the product-state, which does not
contain fluctuations.

The symmetry restoration in the mean-field studies is a major challenge 
to the nuclear structure models. Although, the projection methods which restore
the broken symmetries have been known for more than thirty years,
but have been studied only in simple model cases due to the
numerical difficulties. In most of the realistic studies,
approximate projection methods have been employed \cite{fo96}.
Only recently, the gradient method in the projected-energy surface
has been employed in realistic calculations with Gogny forces \cite{aer.01}.
There has been an unsolved
problem whether a simple HFB like equations can be obtained with
the projected-energy functional. This would allow the usage of
fast numerical algorithms for diagonalisation of matrices as is
done in the ordinary HFB theory. This problem has been recently
addressed and it has been shown that the variation of an arbitrary
energy-functional, which is completely expressed in terms of
the HFB density matrix $\rho$ and the pairing tensor $\kappa$,
results in the HFB like equation \cite{sr2000}. It is noted
that the projected-energy functional can also be expressed in
terms of the HFB density matrices $\rho$ and $\kappa$
and, therefore, one obtains HFB like equations with modified
expressions for the pairing-potential and the Hartree-Fock field.

In the present work, we shall limit ourselves to the restoration
of the gauge symmetry associated with the particle-number projection,
which is important in the discussion of the pairing-correlations. Several
approximate particle-number projection methods have been developed
and the most popular of these is the method of Lipkin and Nogami \cite{ln}.
The purpose of the present work is to perform a systematic study
of these various approximate projection methods and to compare them with the
recently developed number-projected HFB (PHFB) approach. It is shown
that most of the approximate methods including
the Lipkin-Nogami method breakdown in the weak pairing limit. The PHFB method,
on the other hand, gives finite pair-correlations in all the cases
studied.  The paper is organised as follows : a brief survey of the
the particle-number projection methods is given in the next section and
in the appendix. The model Hamitonian used in the present work is expressed in
section III, the numerical results are discussed in section IV and finally the
present work is summarised in section V.

\section{Projection methods}

In this section, we shall briefly describe the approximate particle-number
projection methods used in the literature and the exact number-projected HFB
method proposed recently \cite{sr2000}. Although, the approximate methods are quite
standard and the details can be found, for example, 
in the text book \cite{rs80}. But for  completeness, the essential elements
of these methods are discussed in the following subsections.

\subsection{HFB method}

The many-body Hamiltonian of a system of fermions is
usually expressed in terms of a set of annihilation and creation operators
$(c,c_{}^{\dagger})=(c_{1}^{{}},\ldots,c_{M}^{{}};c_{1}^{\dagger },
    \ldots ,c_{M}^{\dagger })$:
\begin{equation}
H=\sum_{n_1n_2}e_{n_1n_2}^{}c_{n_1}^{\dagger}c_{n_2}^{}+
\frac{1}{4}\sum_{n_1n_2n_3n_4}\overline{v}_{n_1n_2n_3n_4}^{}
c_{n_1}^{\dagger}c_{n_2}^{\dagger}c_{n_4}^{}c_{n_3}^{},
\label{E1}
\end{equation}
where the anti-symmetrized two-body matrix-element is defined by
\begin{equation}
\overline{v}_{n_{1}n_{2}n_{3}n_{4}}^{{}}=\langle
n_{1}n_{2}|V|n_{3}n_{4}-n_{4}n_{3}\rangle .  \label{E2}
\end{equation}
The operators annihilate the bare fermion vacuum
\begin{equation}
c_{n}|-\rangle =0 , \label{E3}
\end{equation}
for all $n$ and provide a basis in the present analysis.

The most basic of the mean-field approaches to describe the 
pairing-correlations is the HFB method. In this method, the
ground-state wavefunction is a product-state
of quasiparticle operators $(\alpha,\alpha_{}^{%
\dagger })= (\alpha _{1}^{},\ldots ,\alpha_{M}^{}; \alpha _{1}^{\dagger
},\ldots ,\alpha _{M}^{\dagger })$. These quasiparticle operators are connected to the
original particle operators by a linear transformation

\begin{eqnarray}
\quad\quad \alpha_{k}^{} &=&\sum_{n}\left( U_{nk}^{\ast
}c_{n}^{{}}+V_{nk}^{\ast}c_{n}^{\dagger }\right) , \label{E4} \\
\quad\quad \alpha_{k}^{\dagger} &=&\sum_{n}\left(
V_{nk}^{{}}c_{n}^{{}}+U_{nk}^{{}}c_{n}^{\dagger }\right) .  \label{E5}
\end{eqnarray}
This transformation can be rewritten in the matrix form as
\begin{equation}
\left(
\begin{array}{c}
\alpha ^{{}} \\
\alpha ^{\dagger }
\end{array}
\right) =\left(
\begin{array}{cc}
U^{\dagger } & V^{\dagger } \\
V^{T} & U^{T}
\end{array}
\right) \left(
\begin{array}{c}
c \\
c^{\dagger }
\end{array}
\right) ={\cal W}^{\dagger }\left(
\begin{array}{c}
c \\
c^{\dagger }
\end{array}
\right) .  \label{E6}
\end{equation}
The quasiparticle operators need to satisfy the same fermion commutation
relations as the original operators in order to preserve the
Fermi-Dirac statistics and, therefore, the transformation matrix is
required to be unitary
\begin{equation}
{\cal W}^{\dagger }{\cal W}={\cal W}{\cal W}^{\dagger }=I,  \label{E7}
\end{equation}
which leads to following relations among the coefficients $U_{nk}$ and $%
V_{nk}$,
\begin{eqnarray}
\quad \quad U^{\dagger }U+V^{\dagger }V=I,\quad \quad &&UU^{\dagger
}+V^{\ast }V^{T}=I,  \label{E8} \\
\quad \quad U^{T}V+V^{T}U=0,\quad \quad &&UV^{\dagger }+V^{\ast }U^{T}=0.
\label{E9}
\end{eqnarray}
The quasiparticle operators annihilate the quasiparticle vacuum $|\Phi
\rangle ,$ defined by
\begin{equation}
\alpha _{k}|\Phi \rangle =0,  \label{E10}
\end{equation}
for all $k$. In mean-field theory, it represents an approximation to the
ground-state of a system and turns out to be a generalized
Slater-determinant \cite{rs80}. Since the quasiparticle transformation
mixes creation and annihilation operators, $|\Phi \rangle $ does not
correspond to a wavefunction with good particle-number. In HFB theory,
we have two kinds of densities, the normal density $\rho$,
and the pairing-tensor $\kappa $, defined as
\begin{equation}
\rho _{nn^{\prime }}~=~\langle \Phi |c_{n^{\prime }}^{\dagger
}c_{n}^{{}}|\Phi \rangle ,\quad \quad \kappa _{nn^{\prime }}~=~\langle \Phi
|c_{n^{\prime }}^{{}}c_{n}^{{}}|\Phi \rangle .  \label{E11}
\end{equation}
These can be expressed in terms of the HFB coefficients as
\begin{equation}
\rho =V^{\ast }V^{T},\quad \quad \kappa =V^{\ast }U^{T}=-UV^{\dagger }.
\label{E12}
\end{equation}
The HFB energy is given by
\begin{equation}
E^{HFB}[\rho,\kappa]  =\frac{ \langle \Phi | H | \Phi \rangle} {
\langle \Phi| \Phi \rangle} = {\rm Tr}\biggr\{ (e + \frac{1}{2}
\Gamma )\rho \biggr\} - \frac{1}{2}
             {\rm Tr} (\Delta \kappa^{\ast}),\label{ehfb}
\end{equation}
where the HFB fields are defined as
\begin{eqnarray}
\Gamma _{n_{1}n_{3}}^{{}} &=&\sum_{n_{2}n_{4}}\overline{v}%
_{n_{1}n_{2}n_{3}n_{4}}^{{}}\rho _{n_{4}n_{2}}^{{}},  \label{E102} \\
\Delta _{n_{1}n_{2}}^{{}} &=&\frac{1}{2}\sum_{n_{3}n_{4}}\overline{v}%
_{n_{1}n_{2}n_{3}n_{4}}^{{}}\kappa _{n_{3}n_{4}}^{{}}.  \label{E103}
\end{eqnarray}
The variation of the HFB energy-functional (\ref{ehfb}) results in
the standard HFB equations :

\begin{equation}\label{hfb}
{\cal H}\left( \begin{array}{c} U\\V \end{array}\right)=E_i \left(
\begin{array}{c} U\\V \end{array}\right), \label{hfbeq1}
\end{equation}
where
\begin{eqnarray}
{\cal H}=&
\left( \begin{array}{cc} e+
\Gamma-\lambda  & \Delta \\
-\Delta^{\ast}  &-(e + \Gamma)^{\ast}+\lambda
\end{array}\right) . \label{hfbeq}
\end{eqnarray}
As already mentioned, the HFB wavefunction does not have a well
defined particle-number and the
Lagrangian parameter $(\lambda)$ is introduced to have the
correct particle-number on the average. The diagonalisation of the HFB matrix
(\ref{hfbeq}) gives rise to the quasiparticle energies ($E_i$) and the excitation
spectrum of the many-body system is constructed from these
quasiparticle energies.

\subsection{Variation after projection - Number-projected (PHFB) equations}

It has been shown recently \cite{sr2000} that the variation after projection
leads to the HFB like equations with modified expressions for the
pairing-field and the Hartree-Fock potential. We have also
obtained these equations using the method of statistical averages and
the derivation is given in the appendix. In this section, we shall present the
relevant expressions of the projected-fields which are used in the
numerical study.
The projected-energy functional is given by \cite{rs80}
\begin{equation}
E^{N}[\rho,\kappa]=\frac{\left\langle \Phi |HP^{N}|\Phi
\right\rangle }{\left\langle \Phi |P^{N}|\Phi \right\rangle
}=\frac{\int d\phi \langle \Phi | {H} e^{i\phi ({\hat N}-N)} |
\Phi \rangle} {\int d\phi \langle \Phi|e^{i\phi ({\hat N}-N)}
|\Phi \rangle}, \label{E26}
\end{equation}
where $\rho$ and $\kappa$ are the unprojected densities given in
Eq. \ref{E12} and the particle-number projection operator is
defined as
\begin{equation}
P^{N}= \frac{1}{2\pi } \int d\phi\ e^{i\phi (\hat{N}-N)}  .  \label{E22}
\end{equation}
It has been demonstrated in ref.\cite{sr2000} that the variation of the
projected-energy results in the HFB equations of the same structure
as in (\ref{hfbeq}) except that the expressions for the fields are
different. The projected HFB equation is given by
\begin{equation}\label{phfb}
{\cal H}^N\left( \begin{array}{c} U\\V \end{array}\right)={\cal E}_i \left(
\begin{array}{c} U\\V \end{array}\right),
\end{equation}
where
\begin{eqnarray}
{\cal H}^N=&
\left( \begin{array}{cc} \varepsilon^N +
\Gamma^N + \Lambda^N   & \Delta^N \\
-{(\Delta^N)}^{\ast}  &-{(\varepsilon^N)}^{\ast} -
{(\Gamma^N)}^\ast - {(\Lambda^N)}^\ast
\end{array}\right).\label{phfbeq}&
\end{eqnarray}
The number-projected expressions for the fields are now given by
\begin{eqnarray}
\varepsilon^N &=&
\frac{1}{2}\int d\phi \,\,y(\phi )\left\{Y(\phi)
{\rm Tr}[e\rho(\phi)] +
[1-2ie^{-i\phi}\sin\phi\rho(\phi)] e C(\phi)\right\} \nonumber \\
&& ~+~h.c.
\label{E201}
\\
\Gamma^N &=&
\frac{1}{2}\int d\phi \,\,y(\phi )\left(Y(\phi)
\frac{1}{2}{\rm Tr}[\Gamma(\phi)\rho(\phi)] +\frac{1}{2}
[1-2ie^{-i\phi}\sin\phi\rho(\phi)]\Gamma(\phi)C(\phi)\right) \nonumber \\
&& ~+~h.c.
\label{E202}
\\
\Lambda^N &=&
-\frac{1}{2}\int d\phi \,\,y(\phi )\left(Y(\phi)
\frac{1}{2}{\rm Tr}[\Delta(\phi)\overline{\kappa}_{}^\ast(\phi)] -
2ie^{-i\phi}\sin\phi\; C(\phi)\Delta(\phi)\overline{\kappa}_{}^{\ast}\right)
\nonumber \\
&& ~+~h.c.
\label{E203}
\end{eqnarray}

\begin{equation}
\Delta^N =\frac{1}{2} \int d\phi\;y(\phi ) e^{-2i\phi }
C\left( \phi \right) \Delta
(\phi )-(..)^{\intercal } \label{E204},
\end{equation}
with
\begin{eqnarray}
\Gamma _{n_{1}n_{3}}^{{}}(\phi )
&=&\sum_{n_2n_4}\overline{v}_{n_1n_2n_3n_4}^{}
\rho_{n_4n_2}^{}(\phi) \label{E501},\\
\Delta _{n_1n_2}^{}(\phi )
&=&\frac{1}{2}\sum_{n_3n_4}\overline{v}_{n_1n_2n_3n_4}^{}
\kappa_{n_3n_4}^{}(\phi) \label{E502},\\
\overline{\Delta}_{n_3n_4}^{\ast}(\phi )
&=&\frac{1}{2}\sum_{n_1n_2}\overline{\kappa}^{\ast}_{n_1n_2}(\phi)
\overline{v}_{n_{1}n_{2}n_{3}n_{4}}^{{}} \label{E503},
\end{eqnarray}

\begin{eqnarray}
\rho (\phi ) &=&C(\phi )\rho, \\
\kappa (\phi ) &=&C(\phi )\kappa =\kappa C_{{}}^{\intercal }(\phi ),\\
\overline{\kappa }(\phi ) &=&e^{2i\phi }\kappa C_{{}}^{\ast }(\phi
)=e^{2i\phi }C_{{}}^{\dagger }(\phi )\kappa,\\
C(\phi ) &=&e^{2i\phi }\left( 1+\rho (e^{2i\phi }-1)\right) ^{-1},\\
x(\phi )&=&\frac{1}{2\pi }\frac{e^{i\phi (N)} \det (e^{i\phi})}
{\sqrt{\det C(\phi )}}, \label{E28} \\
y(\phi) &=&\frac{x(\phi)}{\int dg\,x(\phi)},\;\;\;\int dg\,y(\phi)=1,
\label{E29}
\end{eqnarray}
and
\begin{equation}
Y(\phi )=\,ie^{-i\phi}\sin\phi\;C(\phi )-i
\int d\phi ^{\prime }y(\phi^\prime)e^{-i\phi^\prime}\sin\phi^\prime\;
C(\phi^\prime).
\end{equation}
The term designated by $\Lambda^N$ in Eq. (\ref{E203}) does not appear in the
normal HFB formalism and it can be immediately shown that it vanishes
for the gauge-angle,
$\phi=0$. This term originates from the variation of the pairing-energy with
respect to the density-matrix. In normal HFB theory, 
the pairing-energy depends only
on the pairing-tensor, but the projected HFB pairing-energy also depends
on the density-matrix through the norm-overlap. As a matter of fact,
in a general case, the norm-overlap depends on both 
the density-matrix and the pairing-tensor
(see appendix). But for the special case of number-projection, the term in the
overlap-matrix, depending on the pairing-tensor, can be rewritten
in terms of the density-matrix by using the HFB relation
$(\rho-\rho^2=\kappa \kappa^\dagger)$. Due to this transformation, the
expression for $\Delta^N$ in (\ref{E204}) has a very simple appearance and
reduces to the familiar form in the canonical representation 
\cite{sr2000,dmp64}. 

It should be noted that as compared to the HFB case, the projected
quasiparticle energies obtained from the diagonalisation of the
projected HFB matrix Eq. (\ref{phfbeq}) have no meaning. The only meaningful
quantity is the total projected-energy, which is given by
\begin{eqnarray}
E_{tot}&=&\int d\phi y(\phi) H(\phi), \label{E107} \\
       &=&\int d\phi y(\phi )\left\{ H_{sp}(\phi )+H_{ph}(\phi )+H_{pp}(\phi
)\right\} ,  \label{108}
\end{eqnarray}
where,
\begin{eqnarray}
H_{sp}(\phi ) &=&{\rm Tr}\left( e\rho (\phi )\right) , \\
H_{ph}(\phi ) &=&\frac{1}{2}{\rm Tr}\left( \Gamma (\phi )\rho (\phi )\right),\\
H_{pp}(\phi ) &=&-\frac{1}{2}{\rm Tr}\left( \Delta (\phi )%
\overline{\kappa }_{{}}^{\ast }(\phi )\right).
\end{eqnarray}
In the present work, we shall discuss the pairing-energy ($E_{pair}$) and the
aligned angular-momentum ($<J_x>$) quite extensively, since these give
a measure of the pairing-correlations. The
expressions for these quantities are given by
\begin{eqnarray}
E_{pair}&=&\int d\phi H_{pp}(\phi ) \label{109}, \\
<J_x>   &=&\int d\phi {\rm Tr}\left( J_x\rho (\phi )\right) \label{110}.
\end{eqnarray}

As is clear from Eq. (\ref{phfbeq}), the projected equations have exactly the
same structure as that of normal HFB. Therefore, one can employ the existing
HFB computer codes and only the expressions for the projected-fields need to be
evaluated. The projected-fields now involve the integration over the
gauge-angle, $\phi$. This integration has been performed using the
Gauss-Chebyshev quadrature method \cite{hit79}. In this method,
the integration over the gauge-angle is replaced by a summation.
It can be shown \cite{hit79} that the optimal number of
mesh-points in the summation
which eliminates all the components having undesired particle
numbers is given by
\begin{equation}
 M = { \rm max } \left( {1 \over 2} N, \Omega-{1 \over 2} \right) +1,
\end{equation}
where N is the number of particles and $\Omega$ is the degeneracy of
the model space.

\subsection{Projection after variation (PAV)}

In this projection method, the HFB mean-field solution is used to calculate the
projected quantities. The normal variational calculations are carried out
in the HFB formalism and the solution obtained is then used to
calculate the projected quantities as given by
Eqs. (\ref{108},\ref{109},\ref{110}).
Therefore, this method is closely based on the HFB method and in the event that
HFB solution depicts a transitional behaviour, it would also be reflected in
the PAV results.

\subsection{Lipkin-Nogami (LN) method}

The Lipkin-Nogami (LN) method has been used quite extensively as an
approximation to the particle-number projection approach. In this method, the
HFB energy-function is modified to include the quadratic term
in the Kamlah expansion of the projected-energy \cite{rs80}. The HFB energy is
replaced by
\begin{equation}
E_{HFB} \rightarrow E_{HFB} - \lambda_2 \langle \Delta {\hat N}^2 \rangle ,
\end{equation}
where $ \Delta {\hat N}^2 =  ({\hat N} - \langle {\hat N} \rangle )^2 $
and the Lagrangian parameter $\lambda_2$ is given by \cite{fo96}
\begin{equation}
\lambda_2 = \frac{ \langle \hat H ( \Delta \hat{N}^2 -
\langle \Delta \hat{N}^2 \rangle ) \rangle - \langle {\hat H} \Delta {\hat N}
\rangle \langle \Delta \hat{N}^3 \rangle / \langle \Delta \hat{N}^2 \rangle }
{ \langle \Delta \hat{N}^4 \rangle -  \langle \Delta \hat{N}^2 \rangle^2
- \langle \Delta \hat{N}^3 \rangle^2 /  \langle \Delta \hat{N}^2 \rangle}.
\label{lambda2}
\end{equation}
In the above equations, the expectation value of an operator is calculated with
respect to the HFB vacuum, i.e.,
$\langle \hat O \rangle = \langle \Phi | \hat O | \Phi \rangle $ and can
be calculated by differentiating the norm and the Hamiltonian overlaps
as a function of the gauge-angle, i.e.,
\begin{equation}
\langle \Delta {\hat N}^p \rangle =  2\pi\frac{1}{\imath^p} \frac {\partial^p
 x(\phi)} {\partial \phi^p} |_{\phi=0},
\end{equation}
\begin{equation}
\langle H \Delta {\hat N}^p \rangle =  \frac{1}{\imath^p} \frac {\partial^p
 H(\phi)} {\partial \phi^p} |_{\phi=0},
\end{equation}
where $x(\phi)$ and $H(\phi)$ are defined in Eqs. (\ref{E28}) and (\ref{E107}),
respectively. The expressions for the expectation values required in Eq. (\ref{lambda2})
are given, for example in ref. \cite{fo96}. We have rederived these expectation
values using the expressions (\ref{E28}) and (\ref{E107}), which
essentially involve differentiation of
determinants and the matrix functions. The expectation values are given by
\begin{eqnarray}
\langle \Delta {\hat N}^2 \rangle &=& 2 {\rm Tr} (\chi) ,\\
\langle \Delta {\hat N}^3 \rangle &=& 4 {\rm Tr} (\gamma \chi), \\
\langle \Delta {\hat N}^4 \rangle &=& 3\langle \Delta {\hat N}^2 \rangle^2
+ 8 {\rm Tr} \{\chi(1-6\chi)\},\\
\langle H \Delta {\hat N} \rangle &=& 2 {\rm Tr} (h \chi) - {\cal R}
{\rm Tr} (\Delta \kappa^\ast \gamma),\\
\langle H (\Delta {\hat N}^2 - \langle \Delta {\hat N}^2 \rangle) \rangle
&=& 4 {\rm Tr} [(h \gamma + V_{HF}\{\chi\})\chi ], \nonumber \\
&-&  {\cal R} {\rm Tr} [\Delta \kappa^\ast (1-8\chi)] -
{\rm Tr} [ V_P \{\gamma \kappa \} (\gamma \kappa)^\ast ], \label{hdeln2}
\end{eqnarray}
where,
\begin{eqnarray}
\chi &=& \rho(1-\rho),\\
\gamma &=& 1 - 2\rho ,\\
h    &=& e +\Gamma,\\
V_{HF}\{\chi\}_{ij} &=& \sum_{kl}  \overline{v}_{ikjl} \chi_{lj} ,\\
V_P\{\gamma \kappa \}_{ij} &=& {1 \over 2} \sum_{kl}
\overline{v}_{ijkl} (\gamma \kappa)_{kl}.
\end{eqnarray}
${\cal R}$ in Eq. (\ref{hdeln2}) denotes the real part of the expression.
In the LN method, the calculations proceed exactly as in HFB case except for
the extra $\lambda_2$ term. At each HFB iteration, the $\lambda_2$ is
calculated using the expression (\ref{lambda2}) and this is then used in the
solution of the HFB equations in the next iteration.
\section{Model Hamiltonian}
In the present work, we have carried out extensive numerical
calculations using the model Hamiltonian consisting of a cranked deformed
one-body term, $h^\prime$ and a scalar two-body delta-interaction, $V_2$
\cite{snrp89}. The deformed shell model Hamiltonian employed is given by
\begin{eqnarray}
H&=&h^{\prime}_{def} + V_{2}, \label{E5010}\\
 &=&h_{def} - \omega J_x + V_{2},  \label{E5011}
\end{eqnarray}
where,
\begin{equation}
h_{def}=-4\kappa \sqrt{\frac{4\pi }{5}}\sum_{m}<jm|Y_{20}|jm>c_{jm}^{\dagger
}c_{jm},  \label{E5021}
\end{equation}
and
\begin{equation}
V_{2}={\frac{1}{2}}\sum_{LM}E_{L}^{{}}A_{LM}^{\dagger }A_{LM}^{{}},
\label{E5031}
\end{equation}
with $A_{LM}^{\dagger }=(c_{j}^{\dagger }c_{j}^{\dagger })_{LM}$ and $%
A_{LM}^{{}}=(A_{LM}^{\dagger })^{\dagger }$. For the
antisymmetric-normalized two-body matrix-element ( $E_{J}^{{}}$ ), we use
the delta-interaction which for a single j-shell is given by \cite{gb}
\begin{equation}
E_{L}^{{}}=-G{\frac{(2j+1)^{2}}{2(2L+1)}}\left[
\begin{array}{ccc}
j & j & L \\
\frac{1}{2} & -\frac{1}{2} & 0
\end{array}
\right] ^{2},
\label{E5041}
\end{equation}
where the symbol $[~~]$ denotes the Clebsch-Gordon coefficient.
We use
$G=g\int R^4_{nl}r^2 dr$ as our energy unit and the deformation
energy $\kappa$ is related to the deformation parameter $\beta$.
For the case of $h_{11/2}$ shell, $\kappa$=2.4 approximately
corresponds to $\beta=0.23$ \cite{swi98}.

Although, the present model Hamiltonian is not very realistic and
the results obtained cannot be compared with the experimental data.
Nevertheless, it contains most of the essential components of a
realistic Hamiltonian. The advantage in this model is that it can
be solved exactly and it is possible to test the quality of an
approximate scheme. We consider that it is quite appropriate to
test the number-projection methods in a cranking model, since
with increasing rotational frequency the pairing-correlations
drop and the results become quite sensitive to the projection
method.

\section{Results and discussions}

The detailed numerical calculations have been carried out for six-particles
in $j=11/2$ shell in order to investigate the validity of the
particle-number projection methods presented in section II. The
calculations have been done
with three different pairing-interactions, monopole [L=0 in Eq. (\ref{E5041})],
monopole plus quadupole [L=0 and 2 in Eq. (\ref{E5041})] and with full
delta-interaction to check the sensitivity of
the projection method on the underlying pairing-interaction.

First, we would like to mention
that the projected-HFB calculations almost completely agree with
the exact solutions for all the cases studied and it is not possible to
distinguish them in the figures. In all the figures, we present
only the PHFB results and it should be kept in mind that these results
coincide with the exact solution. In Fig. 1, the results of the
total-energy ($E_{tot}$) and the pairing-energy ($E_{pair}$) are presented
as a function of the strength of the pairing-interaction. For this set of
calculations, we have employed the full delta-interaction. The HFB total-energy
deviates from PHFB with increasing strength of the pairing-interaction. The
results of total-energy with projection after variation appear to be
in good agreement with the results of PHFB. The calculated total-energy
in LN method interpolate between the PHFB and the HFB results. In general,
it is noted from Fig. 1 that the total-energy is fairly reproduced
by all the approximate projection methods.

The critical test of a projection method is how well it reproduces the
pairing-energy. In the lower panel of Fig. 1, the pairing-energy
obtained from different projection methods is presented. The major
problem associated with the HFB theory is that the pairing-energy
shows a phase transition from finite pairing to zero pairing as is clearly
evident from Fig. 1. The HFB pairing-energy is finite for pairing-strength,
$G$ greater than 0.4, but collapses for $G$ less than or equal to 0.4. The PAV
projected-energy has a similar behaviour as that of HFB energy and is expected
since it uses the HFB mean-field solution. It should be noted that the PAV
pairing-energy for certain values of the strength has a higher value than
the HFB pairing-energy. 

The LN pairing-energy is somewhat
between HFB and PHFB pairing-energy and has a non-zero value for
$G=0.4$. But for the lower values of the pairing-strength, the LN pairing-energy
also collapses. Therefore, the LN method which is used so extensively in the
nuclear structure studies appears to be marginally better than the HFB
approach and deviates substantially from the PHFB results. As is evident
from Fig. 1, the PHFB pairing-energy is non-zero for all the
values of the pairing-strength. In the following, we shall now turn to the
rotational aspects of the pairing-energy.

The results of the cranking calculations with monopole-, (monopole plus
quadrupole)- and delta-interactions are presented in Figs. 2, 3 and 4,
respectively. In the upper panel of Fig. 2, a marked difference between the
PHFB and HFB results are obtained for the monopole-interaction at lower
rotational frequencies. The PAV on the other hand gives a good description
for the total-energy and coincides with the PHFB total-energy at lower
rotational frequencies. However, at higher rotational frequencies it shows
a transitional behaviour and goes over to the HFB total-energy. The reason
for this transition is that at higher rotational frequencies, the HFB
pairing-energy vanishes and the results of PAV become identical to HFB.
It should be noted that the mean-field potential, apart from the contribution of the
pairing-interaction ($\Gamma$-term), is fixed in the present model study and
as the pairing-energy vanishes the results of projected-energy are same
as that of unprojected HFB in the PAV method.
The LN total-energy is slightly lower than HFB, but is still far from the
PHFB total-energy. At higher rotational frequencies all the approximate
projection methods coincide with the PHFB, since the pairing-energy becomes
quite weak.

The HFB pairing-energy, shown in the middle panel of Fig. 2, depicts a
transitional behaviour as expected. For $\hbar \omega \leq 0.45$G, the
pairing-energy is finite, but for higher rotational frequencies
it vanishes. This is a major problem with the HFB theory
and is related to the fact that it does not consider the
particle-number fluctuations which become quite important for a
mesoscopic system like an atomic nucleus. The PHFB pairing-energy, on the
other hand, is clearly non-zero for all the rotational frequencies. It does
drop at around $\hbar \omega = 0.55G$ and continues to drop with increasing
rotational frequency, but decreases very smoothly. This smooth drop is
due to the fluctuations which are inherent in the projection formalism. The
PAV pairing-energy is somewhat better at low rotational frequencies, but after the
HFB phase transition is same as that of HFB. The Lipkin-Nogami pairing-energy
shows a slightly smoother drop at the transitional frequency, but then goes to zero
at a slightly higher frequency as compared to HFB.

The observable quantity which is directly related to the pairing-energy
is the aligned angular-momentum along the rotational axes ($<J_x>$).
The nuclei in the ground-state are in a paired phase with maximal
pairing-correlations. However, with increasing rotational frequency
the particles are unpaired and are aligned towards the rotational axes.
The aligned angular-momentum, therefore, is a measure of the unpaired
particles or inversely of pairing-correlations. It is evident from the
bottom panel of Fig. 2 that the HFB aligned angular-momentum shows
a transition from $<J_x> \approx 0$ to $\approx 10$. This jump is related
to the transitional behaviour of the pairing-energy. The results of PAV closely
follow the HFB results. The LN aligned angular-momentum has a smoother
behaviour, but deviates from PHFB. The crossing frequency, which is given
by the point of inflexion in the slope of the $<J_x>$ curve, is same for
HFB and PAV. The crossing for LN is slightly higher than HFB, but is lower than
PHFB by $\hbar \omega = 0.1G$.

The results of the projection methods with
(monopole plus quadrupole)-interaction are presented in Fig. 3. The
overall agreement between
the approximate projection methods ( HFB, PAV and LN ) and the PHFB
results is improved as compared to the results with a pure monopole-interaction.
The results of the total-energy for PAV agree very well with PHFB at lower
rotational frequencies as in the case of monopole interaction. The LN
total-energy is slightly improved as compared to HFB, but deviates considerably
from the PHFB total-energy. The results of the pairing-energy, shown in the
middle panel of Fig. 3, are similar to the results with monopole pairing. The
LN pairing-energy is smoother at the bandcrossing, but vanishes at higher
rotational frequencies. The behaviour of the LN aligned angular-momentum,
shown in the bottom panel of Fig. 3, is now quite similar to $<J_x>$ of PHFB.
In particular, the slope of the curve, which is a measure of the interaction
between the ground-state band and the s-band, is quite similar to that of HFB,
although the bandcrossing is slightly lower than that of PHFB.

The results of the calculations with the full delta-interaction are shown
in Fig. 4. The total-energy obtained from different approximate
particle number-projection methods
is considerably improved as compared to the previous two cases of monopole-
and (monopole plus quadrupole)-interactions. However, the HFB pairing-energy
shows the phase transition from finite pairing to zero pairing as in the
other two cases. The LN pairing-energy on the other hand is quite smooth
and is non-zero up to a very high rotational frequency. Due to this smooth
behaviour, the aligned angular-momentum also shows a smooth behaviour and
has a similar slope as that of PHFB.

%
We have also studied the deformation dependence of the pairing-energy obtained
from various projection methods. As the deformation increases, the pairing
is expected to drop. In particular, for superdeformed shapes the pairing is
quite weak and the results become quite sensitive to the treatment of the
pairing-energy. In Fig. 5, the pairing-energy is presented as a function
of the deformation parameter, $\kappa$ for three different pairing-interactions
shown in the three panels of the figure. As is clear from the figure, the
pairing-energy drops with increasing deformation obtained from all the
methods. However, the HFB pairing-energy goes to zero at around
$\kappa = 5G$ with monopole- and (monopole plus quadrupole)-interactions shown in
the upper two panels of Fig. 5. In the case of full delta-interaction, the
HFB pairing-energy collapses at a slightly higher deformation value of
$\kappa = 5.5$G. The LN pairing-energy has a smooth behaviour as a function
of the deformation, which is quite similar to pairing-energy obtained
in the PHFB case.
However, the LN pairing-energy also goes to zero at higher deformation values,
wheraes the PHFB pairing-energy is finite for all the studied deformation values.

In Fig. 6, the results of the cranking caculations are presented for a
very large deformation value of $\kappa=5.5$ which is analogous to
the deformation of the superdeformed bands. The dynamic moment of inertia
($J^{(2)}$) shown in Fig. 6 is obtianed from the calculated aligned
angular-momentum as a function of the rotational frequency. As is evident
from Fig. 6 the HFB $J^{(2)}$ is quite large as compared to $J^{(2)}$
obtained from PHFB for all the three pairing interactions. The Lipkin-Nogami
$J^{(2)}$ appears to be close to the PHFB $J^{(2)}$ at lower rotational
frequencies. However, at higher rotational frequencies it deviates quite
a lot from the PHFB $J^{(2)}$ and as a matter of fact it becomes
worse than HFB.

\section{Summary}

In the present study, an attempt has been made to test the quality of the
various particle-number projection methods by comparing the
pairing-correlations and other related quantities. First of all, it is
clear from the present study that the mean-field HFB approach appears
to be a better approximation with more realistic pairing-interaction.
This to our knowledge is not discussed in the literature. The HFB results
for total-energy and the aligned angular-momentum appear to be in
better agreement with the PHFB or exact results for the delta-interaction.
This is in contrast to what is expected from the BCS theory. The BCS theory
is known to be a better approximation for the simple monopole-interaction.
This is easy to understand since in BCS theory one only considers the
interaction among the particles in the time-reversed states, which is
appropriate for the
monopole-interaction. But in a more realistic pairing-interaction, one also
has particle-particle correlations which do not not necessarily operate in
time-reversed states. These extra correlations are obviously neglected in
the BCS theory. In the HFB theory, on the other hand, all the particle-particle
correlations are considered in a self-consistent manner. We would also
like to point out that we have considered all the terms arising from
the pairing-interaction. In most of the studies, the exchange term
arising from the pairing-interaction ($\Gamma$) is not considered since
it contributes to the already fitted Hartree-Fock potential. This
contribution turns out to be largest with the delta-interaction. In the
absence of this term, the agreement between HFB and PHFB is
better for the monopole-pairing.

The usefulness of a projection method is evaluated from its description
of the pairing-correlations. The mean-field HFB approach suffers from
a fundamental problem that it leads to a sharp phase transition from
a superfluid to an unpaired phase. This is clear from the results
presented in the present work, the HFB pairing-energy at critical
points goes to zero. The major test of a projection method is how well
it cures for this transitional problem. The PAV method is obviously not
meant to solve this transitional problem, since it uses the HFB solution
to start with. It suffers from the same problems as that of HFB.

The LN method which is used quite extensively appears to improve the
pairing-energy somewhat, but is also seen to breakdown in the weak
pairing limit. It is observed that as a function of the pairing-strength
it does have a non-zero pairing-correlations at $G=0.4$, which is  a
critical point in the HFB case. However, it should be noted that the
pairing-correlations derived from LN are quite weak as compared to
the PHFB case. As a function of the rotational frequency, the LN
pairing-energy has a similar behaviour, it does not approach zero at
the HFB critical frequency, but goes to zero at a slightly
higher frequency. It can be, therefore, concluded from the present study that
Lipkin-Nogami method is marginally better than the HFB approach.

The pairing-energy calculated from the PHFB method is seen to be finite
for all the studied cases. As already mentioned, the calculated
total-energy and the aligned-angular momentum from the PHFB method are
in complete agreement with the exact solutions of the model Hamiltonian.
We consider that the PHFB method is an ideal candidate to
replace the Lipkin-Nogami method in the nuclear structure studies.
It is to be noted that one of the, often stated, advantages of the LN method
is its computational simplicity \cite{mcdn93,gbdfh94,ver96}. The LN
method, apart from the evaluation
of the $\lambda_2$ Lagrangian parameter, is exactly same as that of
HFB approach.
The PHFB equations have also same structure as that of HFB, except that
one needs to evaluate the projected-fields. The evaluation of these
projected-fields requires some extra work. 

We would like to caution that a simple model Hamiltonian 
has been employed
in the present work and the question arises whether the conclusions drawn
here can be extrapolated to a realistic Hamiltonian
and model space. First of all, we would like to mention that the present model
Hamiltonian is not totally unrealistic as it contains most of the
essential elements of a realistic Hamiltonian. The important 
difference is in the
model configuration space. In the present work, we have 
employed the model space of a
single-j shell and a realistic model space contains several major shells.
Due to this major difference, the computation of the projected-fields may take
a longer time. We are presently implementing the PHFB method in a realistic
configuration space and our initial estimate is that the projected 
calculations may increase the numerical work by a factor of three to four as 
compared to the normal HFB case.

\subsubsection*{Acknowledgements}
This work has been supported in part by the Bundesministerium
f\"ur Bildung und Forschung under the project 06 TM 979 and R.R.
acknowledges support from CIC of Argentina.

\section{Appendix}
We shall derive here general expressions for particle-number projected averages
and their variation, valid at zero and finite temperature. The approach is also
applicable to other projections. Let $R$ be an arbitrary many-body density
operator. The average of an operator $O$ with respect to $R$ will be denoted as
\begin{equation}
\langle O\rangle_R={\rm Tr}[OR]\,,\label{A1}
\end{equation}
where the trace is taken in the full grand canonical ensemble.
For a Hamiltonian $H$ satisfying $[H,\hat{N}]=0$, the
average projected-energy is given by
\begin{equation}
E^N=\frac{\langle HP^N\rangle_R}{\langle P^N\rangle_R}=
\int d\phi\, y^N(\phi)H(\phi)\,,\label{A2}
\end{equation}
where
\begin{equation}
y^N(\phi)=\frac{e^{-i\phi N}\langle e^{i\phi \hat{N}}\rangle_R}
{\int d\phi\, e^{-i\phi N}\langle e^{i\phi \hat{N}}\rangle_R},\;\;\;
H(\phi)=\frac{\langle He^{i\phi \hat{N}}\rangle_R}
{\langle e^{i\phi \hat{N}}\rangle_R}\,,\label{A3}
\end{equation}
with $\int d\phi y^N(\phi)=1$. Eq.\ (\ref{A2}) reduces to (\ref{E26}) in the
pure limit $R=|\Phi\rangle\langle\Phi|$.

Let us consider now a density of the form
\begin{equation}
R=e^{-\beta K}/{\rm Tr}[e^{-\beta K}]\,\label{A4}\,,
\end{equation}
where $K$ is an hermitian {\it one-body} operator. In a fermion system, the
most general one-body $K$ can be written as
\begin{eqnarray}
K&=&\sum_{i,j}{\cal K}^{11}_{ij}c^\dagger_ic_j+
{\textstyle\frac{1}{2}}({\cal K}^{20}c^\dagger_ic^\dagger_j
+{\cal K}^{02}_{ij}c_ic_j)=k_0+{\textstyle\frac{1}{2}}Z^{\dagger}{\cal K}Z\,,
\label{A5}\\
Z&=&\left(\begin{array}{c}c\\c^\dagger\end{array}\right),\;\;\;
Z^\dagger=(c^\dagger,c),\;\;\;\;
{\cal K}=\left(\begin{array}{cc}{\cal K}^{11} &
{\cal K}^{20}\\ {\cal K}^{02} & -({\cal K}^{11})^{T}\\
\end{array}\right)\,,\nonumber
\end{eqnarray}
with $k_0=\frac{1}{2}{\rm Tr}\,{\cal K}^{11}$ and ${\cal K}^{20}$,
${\cal K}^{02}$ antisymmetric matrices. For $K$ hermitian,
${\cal K}^\dagger={\cal K}$, i.e., $({\cal K}^{11})^{T}=({\cal K}^{11})^*$,
${\cal K}^{02}=-({\cal K}^{20})^*=({\cal K}^{20})^\dagger$.
The norm in (\ref{A4}) is
\begin{equation}
{\rm Tr}[\,e^{-\beta K}]=e^{-\beta k_0}\det[1+e^{-\beta{\cal K}}]^{1/2}\,.
\label{A6}
\end{equation}
It is convenient to define a generalized density matrix \cite{rs80}, of
elements ${\cal R}_{mn}=\langle Z^\dagger_nZ_m\rangle_R$, that comprises
both the standard density matrix and the pairing tensors and is a function
of ${\cal K}$:
\begin{equation}
{\cal R}\equiv\left(\begin{array}{cc}\rho&\kappa\\\tilde{\kappa}&
\;1-\rho^{T}\end{array}\right)=[1+e^{\beta{\cal K}}]^{-1},\label{A7}
\end{equation}
where $\rho_{ij}=\langle c^\dagger_jc_i\rangle_R$,
$\kappa_{ij}=\langle c_jc_i\rangle_{R}$,
$\tilde{\kappa}_{ij}=\langle c^\dagger_jc^\dagger_i\rangle_R$,
For $K$ hermitian, ${\cal R}^\dagger={\cal R}$, i.e., $\rho^{T}=\rho^*$,
$\tilde{\kappa}=-\kappa^*=\kappa^{\dagger}$. {\it All} averages with respect
to $R$ can be expressed in terms of ${\cal R}$. In particular, for an
operator $O$ of the form (\ref{A5}),
\[\langle O\rangle_R={\rm Tr}[{\cal O}^{11}\rho+
{\textstyle\frac{1}{2}}({\cal O}^{20}\tilde{\kappa}+{\cal O}^{02}\kappa)]=
o_0+{\textstyle\frac{1}{2}}{\rm Tr}[{\cal O}{\cal R}]\,.\]
Let us write now the particle-number operator $\hat{N}=\sum_ic^\dagger_ic_i$ as
\begin{equation}
\hat{N}=n_{0}+{\textstyle\frac{1}{2}} Z^\dagger{\cal N}Z\,,\;\;\;\;
{\cal N}=\left(\begin{array}{cc}1 & \;0\\ 0 & \;-1\end{array}\right)\,,
\end{equation}
with $2n_0$ the dimension of the single particle space. As the product of
exponentials of one-body operators is the exponential of a one-body-operator,
we have  $e^{i\phi\hat{N}}e^{-\beta K}=e^{-\beta K(\phi)}$, with $K(\phi)$ a
{\it one-body} operator determined by \cite{rr94}
\begin{eqnarray}
K(\phi)&=&k_0(\phi)+{\textstyle\frac{1}{2}}Z^\dagger{\cal K}(\phi)Z\,,
\label{A81}\\
e^{-\beta{\cal K}(\phi)}&=&e^{i\phi {\cal N}}e^{-\beta{\cal K}}=
e^{i\phi{\cal N}}{\cal R}(1-{\cal R})^{-1}\,,\label{A82}
\end{eqnarray}
and $k_0(\phi)=k_0-i\phi n_0/\beta$. Eqs.\ (\ref{A4})--(\ref{A82}) lead then to
\begin{eqnarray}
\langle e^{i\phi\hat{N}}\rangle_R&=&e^{i\phi n_0}
\frac{\det[1+e^{-\beta{\cal K}(\phi)}]^{1/2}}
{\det[1+e^{-\beta{\cal K}}]^{1/2}}=\det[e^{i\phi/2}{\cal C}(\phi)]^{1/2}\,,\\
{\cal C}(\phi)&=&1+(e^{i\phi {\cal N}}-1){\cal R}\,,
\end{eqnarray}
which enables the evaluation of $y^N(\phi)$. Besides, by means of Wick's
theorem, $H(\phi)$ can be expressed in terms of the elements of the
non-hermitian density matrix determined by
$R(\phi)=e^{i\phi \hat{N}}R/\langle e^{i\phi \hat{N}}\rangle_R
=e^{-\beta K(\phi)}/{\rm Tr}[e^{-\beta K(\phi)}]$,
\begin{equation}
{\cal R}(\phi)\equiv
\left(\begin{array}{cc}\rho(\phi)&\kappa(\phi)\\\tilde{\kappa}(\phi)&
\;1-\rho^{T}(\phi)\end{array}\right)
=[1+e^{\beta{\cal K}(\phi)}]^{-1}
=e^{i\phi {\cal N}}{\cal R}{\cal C}^{-1}(\phi)\,,
\end{equation}
where $\rho(\phi)=\langle c^\dagger_jc_i\rangle_{R(\phi)}$,
$\kappa(\phi)=\langle c_jc_i\rangle_{R(\phi)}$,
$\tilde{\kappa}(\phi)=\langle c^\dagger_jc^\dagger_i\rangle_{R(\phi)}$.
In the case (\ref{E1}),
\begin{equation}
H(\phi)={\rm Tr}\{[e+{\textstyle\frac{1}{2}}\Gamma(\phi)]\rho(\phi)+
{\textstyle\frac{1}{4}}[\Delta(\phi)\tilde{\kappa}(\phi)+
\tilde{\Delta}(\phi)\kappa(\phi)]\}\,,
\label{A9}
\end{equation}
with $\Gamma_{ij}(\phi)=\sum_{k,l}\bar{v}_{ikjl}\rho_{lk}(\phi)$,
$\Delta_{ij}(\phi)=\frac{1}{2}\sum_{k,l}\bar{v}_{ijkl}\kappa_{kl}(\phi)$,
$\tilde{\Delta}_{ij}(\phi) = \frac{1}{2}\sum_{k,l}\bar{v}_{klij}
\tilde{\kappa}_{kl}(\phi)$.
Let us now evaluate the variation of (\ref{A2}) with respect to ${\cal R}$.
We have
\begin{eqnarray}
\delta \langle e^{i\phi \hat{N}}\rangle_R&=&{\textstyle\frac{1}{2}}
\langle e^{i\phi\hat{N}}\rangle_R\,{\rm Tr}
\{{\cal C}^{-1}(\phi)[e^{i\phi {\cal N}}-1]\delta{\cal R}\}\,,\nonumber\\
\delta{\cal R}(\phi)&=&e^{i\phi {\cal N}}\tilde{\cal C}^{-1}(\phi)
(\delta{\cal R}){\cal C}^{-1}(\phi),
\;\;\;\tilde{\cal C}(\phi)=1+{\cal R}(e^{i\phi {\cal N}}-1)\,,\nonumber\\
\delta H(\phi)&=&{\textstyle\frac{1}{2}}{\rm Tr}[{\cal H}(\phi)
\delta{\cal R}(\phi)]={\textstyle\frac{1}{2}}{\rm Tr}[{\cal C}^{-1}(\phi)
{\cal H}(\phi)e^{i\phi {\cal N}}\tilde{\cal C}^{-1}(\phi)\delta{\cal R}]
\,,\nonumber
\end{eqnarray}
where ${\cal H}(\phi)$ is defined by ${\frac{1}{2}}{\cal H}_{mn}(\phi)=
\partial H(\phi)/\partial{\cal R}_{nm}(\phi)$, i.e.,
\begin{eqnarray}
{\cal H}(\phi)&=&\left(\begin{array}{cc}h(\phi)&\Delta(\phi)\\
\tilde{\Delta}(\phi) & -h^{T}(\phi)\end{array}\right)\,,\label{A10}\\
h_{ij}(\phi)&=&\frac{\partial H(\phi)}{\partial\rho_{ji}(\phi)}\,,
\;\;{\textstyle\frac{1}{2}}\Delta_{ij}(\phi)=
\frac{\partial H(\phi)}{\partial\tilde{\kappa}_{ji}(\phi)}\,,
\;\;{\textstyle\frac{1}{2}}\tilde{\Delta}_{ij}(\phi)=
\frac{\partial H(\phi)}{\partial \kappa_{ji}(\phi)}\,.
\label{A11}
\end{eqnarray}
Eqs.\ (\ref{A10})--(\ref{A11}) are valid for {\it any} many-body Hamiltonian.
In the case (\ref{A9}), $h(\phi)=e+\Gamma(\phi)$ and $\Delta(\phi)$,
$\tilde{\Delta}(\phi)$ coincide with the previous definitions.
We finally obtain
\begin{equation}
\delta E^N={\textstyle\frac{1}{2}}{\rm Tr}[{\cal H}^N\delta{\cal R}]\,,
\label{A12}
\end{equation}
where ${\cal H}^N$, defined by ${\frac{1}{2}}{\cal H}^N_{mn}=
\partial E^N/\partial{\cal R}_{nm}$, is given by
\begin{eqnarray}
{\cal H}^N&=&\int d\phi\, y^N(\phi){\cal C}^{-1}(\phi)[(e^{i\phi{\cal N}}-1)
(H(\phi)-E^N)+{\cal H}(\phi)e^{i\phi {\cal N}}\tilde{\cal C}^{-1}(\phi)]
\nonumber\\&=&\left(\begin{array}{cc}h^N & \Delta^N\\ \tilde{\Delta}^N &
-(h^N)^{T}\end{array}\right)\,.\label{A13}
\end{eqnarray}
For $H$ hermitian, $({\cal H}^N)^\dagger={\cal H}^N$.
The first term ${\cal C}^{-1}(\phi)(e^{i\phi{\cal N}}-1)(H(\phi)-E^N)$ arises
from the variation of $y^N(\phi)$.

Eq.\ (\ref{A12}) holds for {\it arbitrary} variations $\delta{\cal R}$.
For {\it unitary} variations  $\delta{\cal R}=i[\delta{\cal W},{\cal R}]$,
Eq.\ (\ref{A12}) leads to
$\delta E^N=-{\frac{1}{2}}i{\rm Tr}\{[{\cal H}^N,{\cal R}]\,\delta{\cal W}\}$.
Stability against these variations leads then to the necessary condition
\begin{equation}
[{\cal H}^N,{\cal R}]=0\,,
\end{equation}
which has the same form as the standard HFB condition \cite{rs80} except that
the HFB matrix ${\cal H}$ has been replaced by ${\cal H}^N$.

All previous expressions hold actually in any basis of single particle or
quasiparticle states, after appropriate transformation of ${\cal N}$. The
matrix ${\cal K}$ can be diagonalized with a transformation (\ref{E6}), i.e.,
$Z'={\cal W}^\dagger Z$, with $(Z')^\dagger=(\alpha^\dagger,\alpha)$.
We may then write ${\cal K}={\cal W}{\cal K}'{\cal W}^\dagger$, with
${\cal K}'$ diagonal, and
\begin{equation}
{\cal R}={\cal W}R'{\cal W}^\dagger,\;\;\;{\cal R}'=
\left(\begin{array}{cc}f&0\\0& \;1-f\end{array}\right)\,,
\end{equation}
with $f_k=\langle \alpha_k^\dagger\alpha_k\rangle_R=(1+e^{\beta E_k})^{-1}$
and $E_k$ the positive eigenvalues of ${\cal K}$.

{\it  $T=0$ limit}. For $T=\beta^{-1}\rightarrow 0$, $f_k\rightarrow 0$
(assuming $E_k\neq 0$) and $R\rightarrow|0\rangle\langle 0|$, with
$|0\rangle$ the vacuum of the quasiparticles $\alpha_k$. In this limit,
\begin{equation}
{\cal R}'\rightarrow\left(\begin{array}{cc}0&0\\0&1\end{array}\right)\,,
\;\;\;{\cal R}\rightarrow\left(\begin{array}{cc}VV^\dagger & \;VU^\dagger\\
UV^\dagger &\;UU^\dagger \end{array}\right)^*\,,
\label{A14}
\end{equation}
implying $\rho=(VV^\dagger)^*$, $\kappa=(VU^\dagger)^*=\tilde{\kappa}^\dagger$,
and ${\cal R}^2={\cal R}$, i.e., $\rho^2+\kappa\tilde{\kappa}=\rho$,
$\rho\kappa=\kappa\rho^{T}$. Using these relations, we obtain the simplified
expressions
\begin{eqnarray}
\det[e^{i\phi/2}{\cal C}(\phi)]^{1/2}&\rightarrow&\det[e^{i\phi}D(\phi)]^{1/2}
\,,\;\;\;D(\phi)=e^{i\phi}\rho+e^{-i\phi}(1-\rho)\,,\nonumber\\
{\cal R}(\phi)&\rightarrow&\left(\begin{array}{cc} e^{i\phi}D^{-1}(\phi)\rho &
e^{i\phi}D^{-1}(\phi)\kappa\\ e^{-i\phi}\tilde{\kappa}D^{-1}(\phi) &
\;1-e^{i\phi}[D^{-1}(\phi)\rho]^T\end{array}\right)\,,\label{A15}
\end{eqnarray}
with ${\cal C}^{-1}(\phi)=1+(e^{-i\phi{\cal N}}-1){\cal R}(\phi)$,
$\tilde{\cal C}^{-1}(\phi)=1-e^{-i\phi{\cal N}}{\cal R}(\phi)
(e^{i\phi{\cal N}}-1)$. With the replacements $C(\phi)=e^{i\phi}D^{-1}(\phi)$,
$\bar{\kappa}(\phi)=\tilde{\kappa}^\dagger(\phi)$,
$\bar{\Delta}(\phi)=\tilde{\Delta}^\dagger(\phi)$, Eq.\ (\ref{A15}) leads to
Eqs.\ (\ref{E501})--(\ref{E503}).

For $T=0$ variations that preserve the idempotent condition
${\cal R}^2={\cal R}$, we may employ a simplified matrix ${\cal H}^N$, which
can be directly derived from the above expressions. Its elements are
\begin{eqnarray}
h^N&=&\int d\phi\, y^N(\phi) D^{-1}(\phi)\{i\sin\phi(H(\phi)-E^N)\nonumber\\
&&+[h(\phi)-i\sin\phi(e^{i\phi}
\kappa\tilde{\Delta}(\phi)+e^{-i\phi}\Delta(\phi)\tilde{\kappa})]
D^{-1}(\phi)\}\,,\\
\Delta^N&=&\int d\phi\,y^N(\phi)e^{-i\phi}D^{-1}(\phi)\Delta(\phi)\,,\;\;\;
\tilde{\Delta}^N=\int d\phi\,y^N(\phi)e^{i\phi}\tilde{\Delta}(\phi)D^{-1}(\phi)
\,.\nonumber
\end{eqnarray}
In the case of Hamiltonian (\ref{E1}), with the previous replacements these
expressions lead to Eqs.\ (\ref{E201})--(\ref{E204}).


\begin{figure}
\caption{
The results of the total energy $(E_{tot})$ and the
pair-energy $(E_{pair})$ as a function of the strength of the
pairing-interaction $(G)$ using various particle-number projection
methods.
}
\label{figure.1}
\end{figure}

\begin{figure}
\caption{
The results of the total energy $(E_{tot})$, the
pair-energy
$(E_{pair})$ and the alignment $(J_x)$ for six-particles
in a deformed $j=11/2$ orbitial using the monopole
interaction. The results are obtained with
Hartree-Fock-Bogoliubov (HFB), projection after variation (PAV),
Lipkin-Nogami (LN) and projected HFB (PHFB) approaches.
}
\label{figure.2}
\end{figure}

\begin{figure}
\caption{
The results of the total energy $(E_{tot})$, the
pair-energy
$(E_{pair})$ and the alignment $(J_x)$ using the monopole plus quadrupole
pairing interaction.
}
\label{figure.3}
\end{figure}

\begin{figure}
\caption{
The results of the total energy $(E_{tot})$, the
pair-energy
$(E_{pair})$ and the alignment $(J_x)$ using the full delta-
interaction.
}
\label{figure.4}
\end{figure}

\begin{figure}
\caption{
The results of the pair-energy
$(E_{pair})$ as a function of the deformation parameter ($\kappa$). (a) gives
the results with monopole-pairing, (b) presents the results with (monopole plus
quadrupole)-interaction and (c) gives the results with full delta-interaction.
}
\label{figure.5}
\end{figure}

\begin{figure}
\caption{
The dynamic moment of ineratia for $\kappa=5.5$. (a) gives
the results with monopole-pairing, (b) presents the results with (monopole plus
quadrupole)-interaction and (c) gives the results with full delta-interaction.
}
\label{figure.6}
\end{figure}

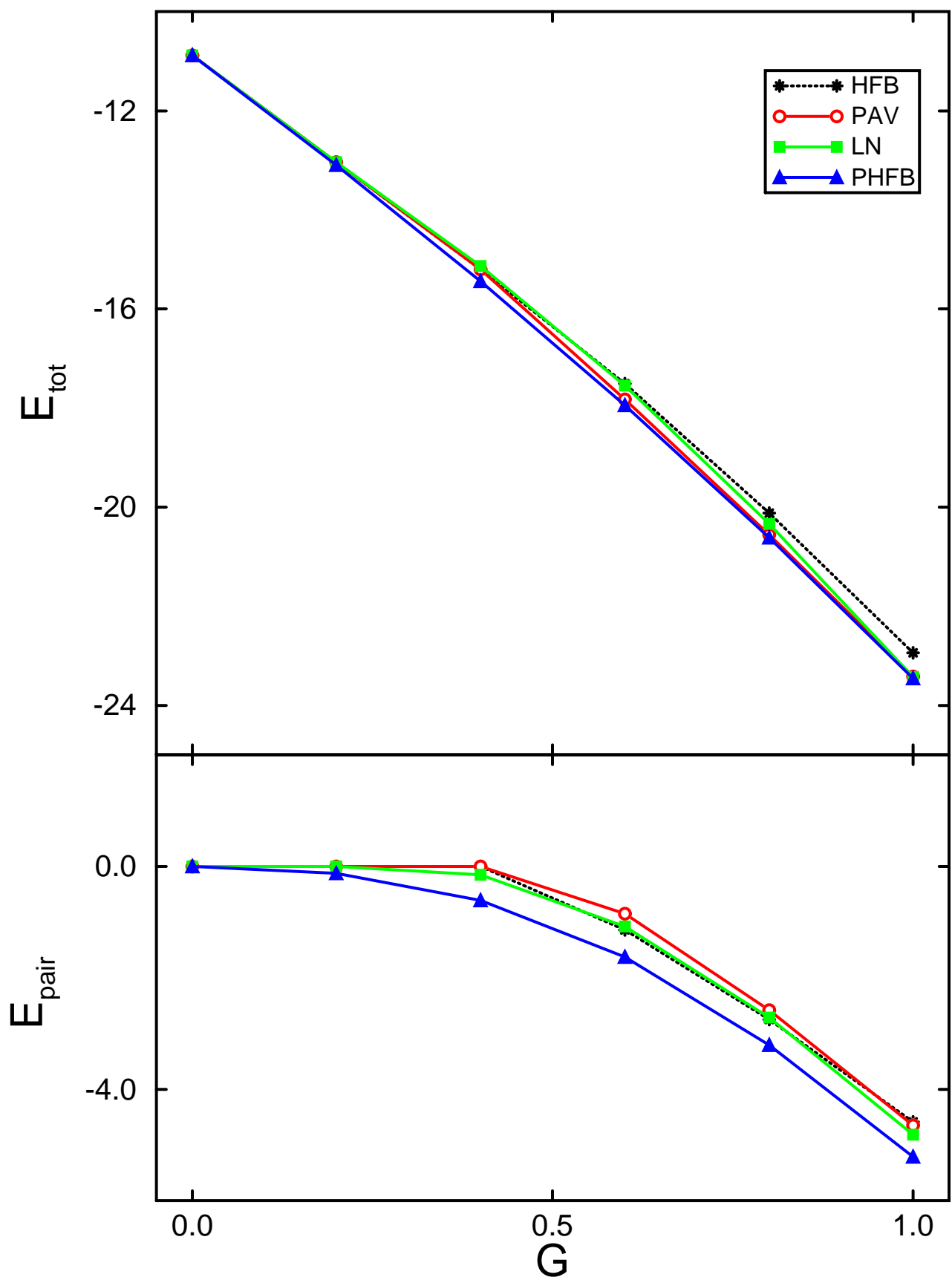
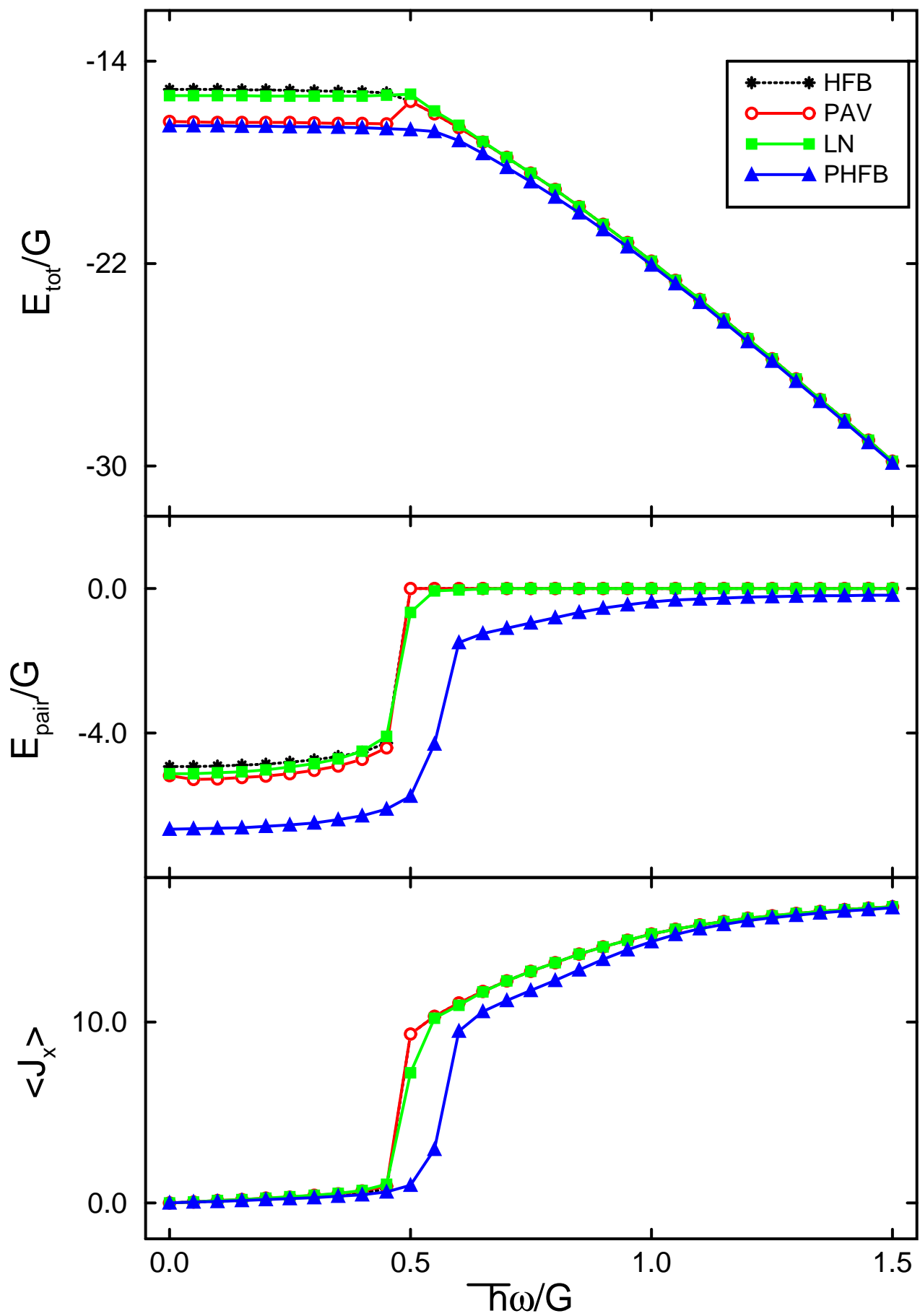
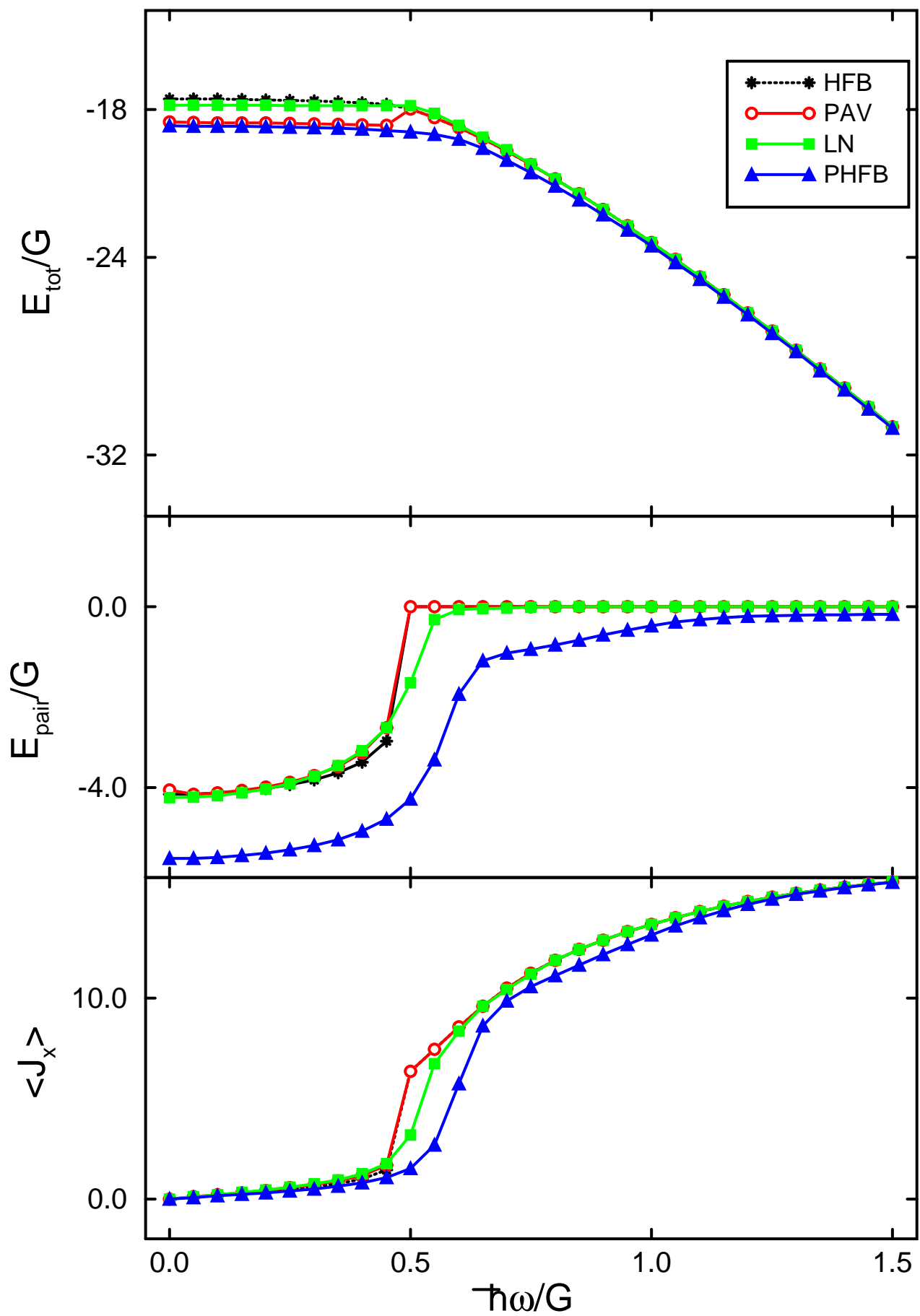
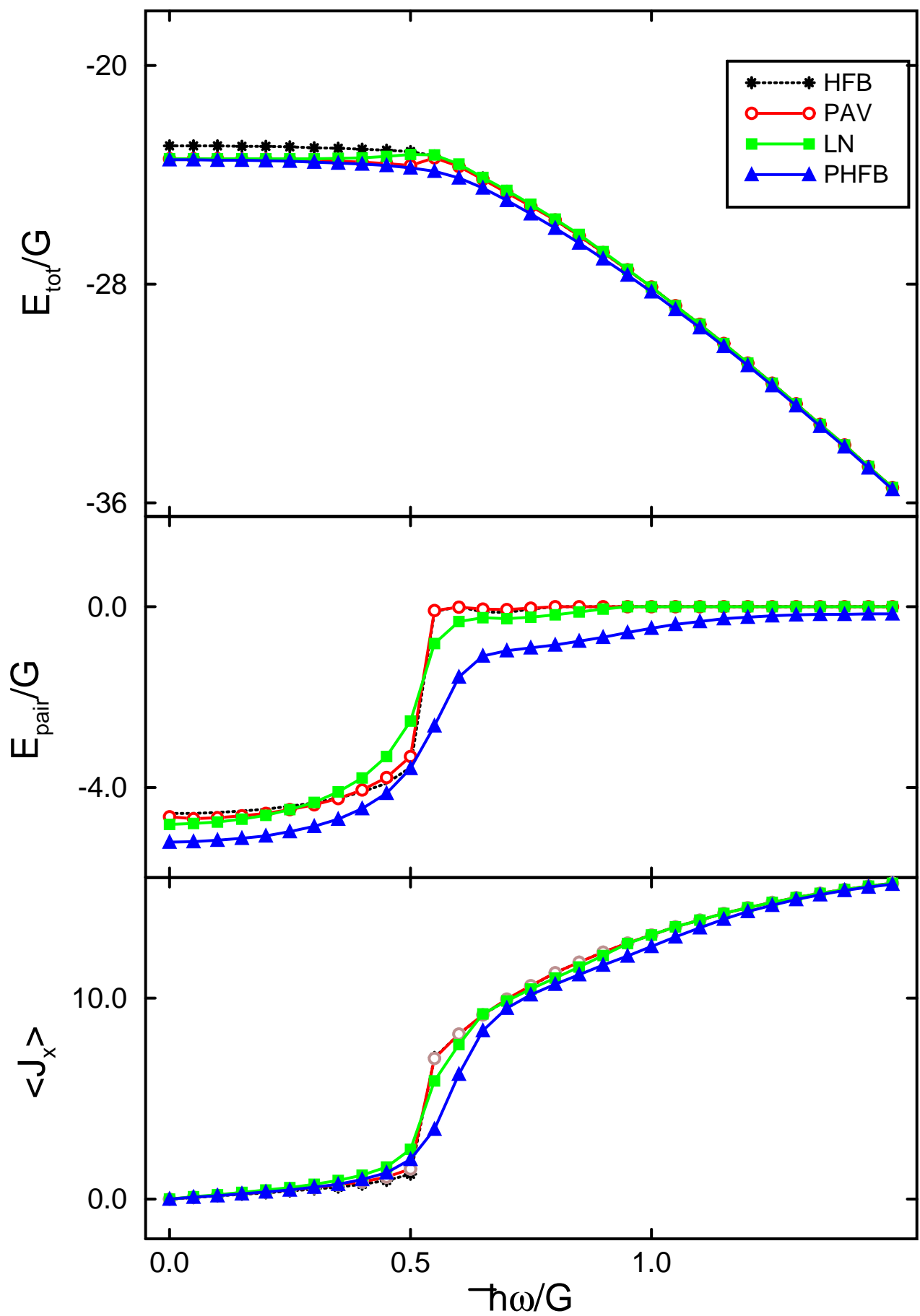
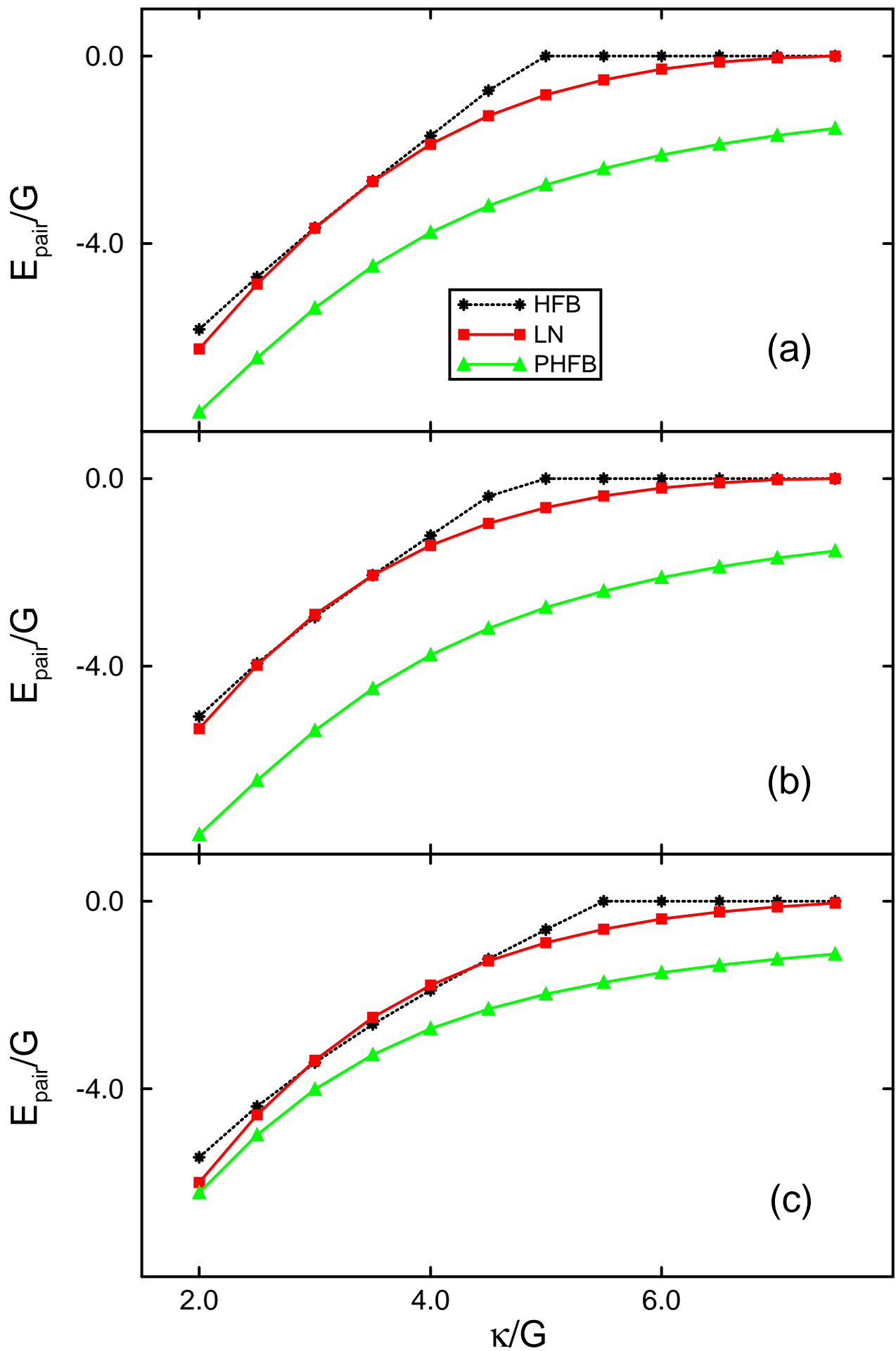
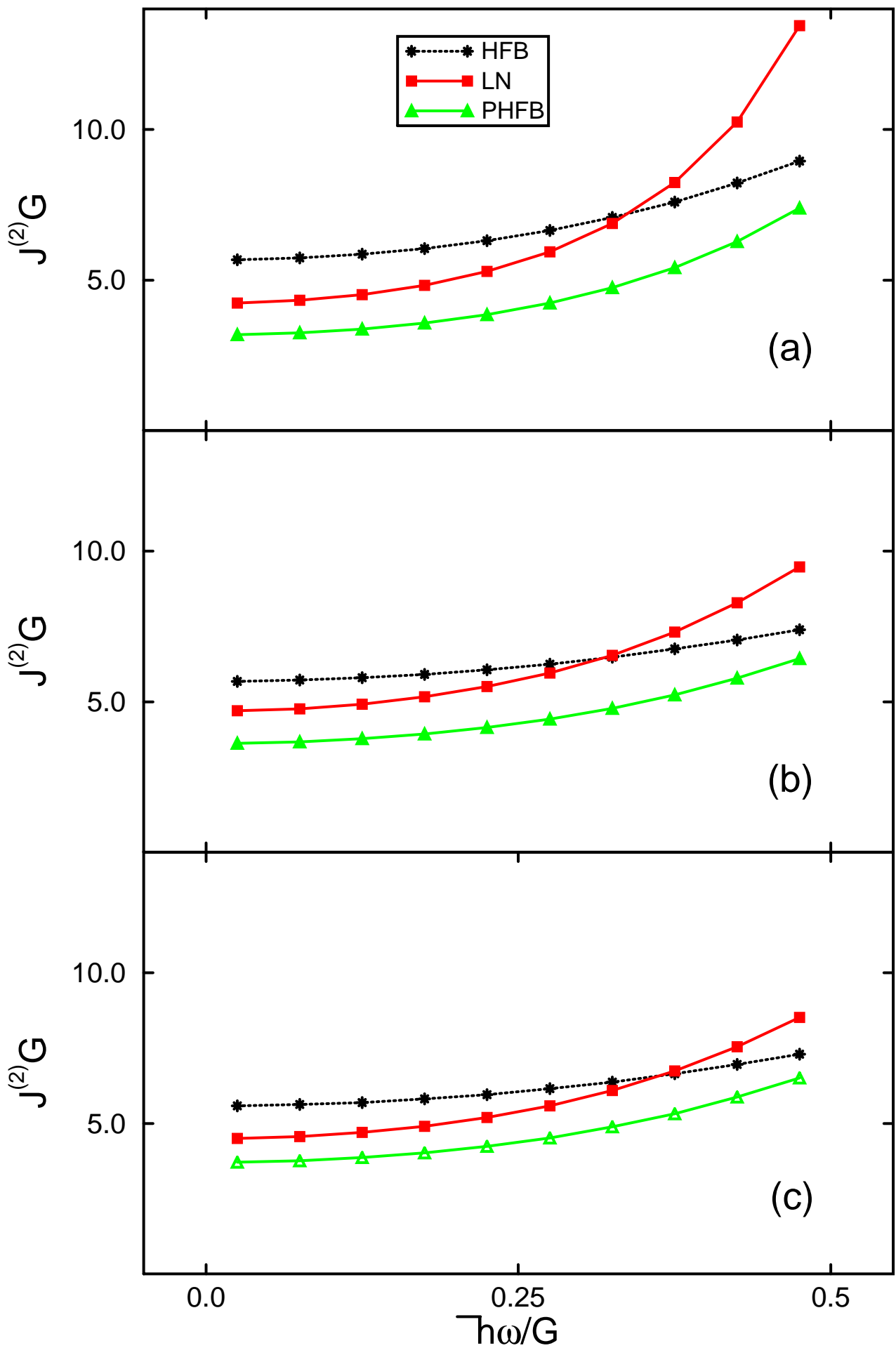
\end{document}